# The inviscid Burgers equation with fractional Brownian initial data: the dimension of regular Lagrangian points

## G. Molchan

Institute of Earthquake Prediction Theory and Mathematical Geophysics,

Russian Academy of Science, 84/32 Profsoyuznaya st.,

117997, Moscow, Russian Federation

E-mail address: molchan@mitp.ru

*Abstract*. Fractional Brownian motion, H-FBM, of index 0<H<1 is considered as initial velocity in the inviscid Burgers equation It is shown that the Hausdorff dimension of regular Lagrangian points at any moment t is equal to H. This fact validates the Sinai-Frisch conjecture known since 1992.

## 1. Introduction.

Ya. Sinai and U. Frisch initiated in 1992 the study of fractal properties of solutions of the inviscid Burgers equation with a stochastic initial velocity $u_0(x)$. One of the problems that arose was the Hausdorff dimension of regular Lagrangian points $S$. These points describe the initial locations of those fluid particles which have not collided until a fixed time. The original model of $u_0(x)$ was Fractional Brownian motion, H-FBM, with Hurst parameter 0<H<1.

Sinai [12] showed that $\dim S = 1/2$ for the Brownian motion case (H=1/2). Handa [7] found simple arguments to derive a lower bound of $\dim S$, namely, $\dim S \geq H$. The exact equality $\dim S = H$ is known as a conjecture since 1992, [12, 13]. Molchan and Khokhlov [11] showed that the validity of $\dim S = H$ can be formulated in terms of the persistence probability:

$$p_T(I_H, \Delta_T) = P\{I_H(x) := \int_0^x w_H(s)ds \leq 1, x \in \Delta_T\}, \qquad (1)$$

namely, $\dim S = H$ if $\Delta_T = (-T, T)$ and

$$\theta[I_H, \Delta_T] := \lim_{T \to \infty} (\log 1 / p_T(I_H, \Delta_T)) / \log T \geq 1 - H \quad . \qquad (2)$$

It is generally a very difficult problem to derive exact values of the persistent exponents $\theta[\xi, \Delta_T]$ for non-markovian processes $\xi$. The current state of the problem can be found in the surveys [3,4]. In addition, two recent works are important in this context.

Aurzada et al. [2] consider the persistence probabilities for stochastic sequences with stationary increments in $\Delta_T = (0,T)$. This approach has considerably strengthened and simplified the proof of Molchan's result [9], viz.,

$$\theta[w_H,(0,T)] = 1 - H.$$

Dembo et al. [5] considered the sequence $I(k) = \sum_1^k \xi_i$, where the $\{\xi_i - \xi_{i-1}\}$ are independent, identically distributed random variables with zero mean. In the case of squared integrability of $\xi_i - \xi_{i-1}$, we have $\theta[I(\cdot),(1,N)] = 1/4$. This is exactly the persistence exponent in the original Sinai problem [12] for the integrated Brownian motion on $\Delta_T = (0,T)$. Note that the situation with H-FBM is more complicated. There are only some theoretical and numerical arguments in favor of the conjecture $\theta[I_H,(0,T)] = H(1-H)$, [10].

Our goal is to show that $\theta[I_H,(-T,T)] \geq 1 - H$, thereby proving that $\dim S = H$.

## 2. The problem

We consider the inviscid Burgers equation
$$\partial_t u + u \partial_x u = \nu \partial_x^2 u, \quad \nu \downarrow 0,$$
$$u(0,x) = u_0(x),$$
where $u_0(x)$ is continuous and $U(x) = \int^x u_0(s)ds = o(x^2)$ as $x \to \infty$. Roughly speaking, this equation applies to the dynamics of completely inelastic particles on $R^1$, [13,15]. Each infinitesimal particle located at $x$ has a mass $dx$ and an initial moment $dU(x)$. On colliding, the particles coalesce and continue movement following the conservation laws of mass and momentum. Initial positions of those particles that have not collided until time $t_0 = 1$ make up the set of regular Lagrangian points $S$.

Formally, $S$ is the topological support of the measure $dC'(x)$, where $C(x)$ is a convex minorant of $U(x) + x^2/2$. Our problem is the Hausdorff dimension of $S$ for the case where $u_0(x)$ is fractional Brownian motion.

H-FBM is a centered Gaussian random process $w_H(x)$ with correlation function
$$Ew_H(x)w_H(y) = 0.5(|x|^{2H} + |y|^{2H} - |x-y|^{2H}).$$

H-FBM is H-self-similar and has stationary increments, i.e., the following relation

$$\{w_H(\lambda x + x_0) - w_H(x_0)\} \doteq \{\lambda^H w_H(x)\} \tag{3}$$

holds in the sense of the equality of finite-dimensional distributions for any fixed $x_0$ and $\lambda > 0$.

By (3), the integrated fractional Brownian motion $I_H(x) := \int_0^x w_H(s)ds$ has a similar property:

$$I_H(x + x_0) - I_H(x_0) - w_H(x_0)x \doteq I_H(x). \tag{4}$$

**Proposition 1** (Handa, [7]). If $u_0(x) = w_H(x)$, and S is the set of regular Lagrangian points, then $\dim S \geq H$.

We remind the elegant Handa's arguments. At points $x_i \in S$, the curves $U(x) + x^2/2$ and $C(x)$ are tangent to each other. Therefore,

$$C'(x_1) - C'(x_2) = w_H(x_1) - w_H(x_2) + x_1 - x_2.$$

It is well known that H-FBM is $\gamma$ - Holder continuous with any $\gamma < H < 1$ (see e.g., [14]). Consequently,

$$|C'(x_1) - C'(x_2)| < K_\gamma |x_1 - x_2|^\gamma$$

holds on any fixed unit interval with some random constant $K_\gamma$. Hence, by the Frostman lemma [6], we get the desired result: $\dim S \geq \gamma = H - \varepsilon$ for any $\varepsilon > 0$.

**Proposition 2.** (Molchan&Khokhlov [11]). Under conditions of Proposition 1, $\dim S \leq \gamma < 1$ holds, if one of the following persistence probabilities $p_T$:

$$P\{I_H(x) \leq 1, x \in (-T,T)\} \quad \text{or} \quad P\{I_H(x) \leq 0, |x| \in (1,T)\},$$

does not exceed $T^{-(1-\gamma)+\varepsilon}$ for any $\varepsilon > 0$ as $T \to \infty$.

. **Proposition 3.** For any persistence probability $p_T$ from Proposition 2 the following holds:

$$\log p_T / \log T \leq -(1-H) + c(1/\sqrt{\log T}),$$

**Corollary.** The Hausdorff dimension of regular Lagrangian points in the inviscid Burgers equation with H-FBM initial data is equal to H.

## 3. Proof of proposition 3.

Consider the sequence $I_T = \{I_H(k), k = 0,1...N \subset [0,T]\}$ and its convex majorant $I_T(x)$. The majorant is a piecewise linear function with the nodal points $\{k_i\}$ and slopes at these points: $\gamma_{k_i}^-$ (left-hand) and $\gamma_{k_i}^+$ (right-hand).

If $k_i = k$, then

$$\gamma_k^- = \min_{1 \le p \le k}(I_H(k) - I_H(k-p))/p, \tag{5}$$

$$\gamma_k^+ = \max_{1 \le p \le N-k}(I_H(k+p) - I_H(k))/p, \tag{6}$$

$$\gamma_k^- \ge \gamma_k^+. \tag{7}$$

Consider the functional

$$F = \sum_{k=1}^{N-1}[\gamma_k^- - \gamma_k^+]_+, \tag{8}$$

where $[x]_+ = x \cdot 1_{x \ge 0}$. By (5-7), the k-th term in F is non-zero if and only if the following event, $A_k$, takes place:

$$A_k = \begin{cases} I_H(k-p) \le I_H(k) - p\gamma_k^-, p \in (1,...,k)...... \\ I_H(k+p) \le I_H(k) + p\gamma_k^+, p \in (1,...,N-k) \\ \gamma_k^- \ge \gamma_k^+ ......................................................... \end{cases} \tag{9}$$

The event $A_k$ means that k is the nodal point of $I_T(x)$. But then

$$F = \sum_i [\gamma_{k_i}^- - \gamma_{k_i}^+]_+ = \sum_i \gamma_{k_i}^- - \gamma_{k_{i+1}}^- = \gamma_0^+ - \gamma_N^-$$

$$= \max_{1 \le p \le N}[I_H(p)/p] - \min_{1 \le p \le N}[(I_H(N) - I_H(N-p))/p],$$

*The mean value of F.* Because $Ew_H(N) = 0$, we have

$$E(-\gamma_N^-) = E\{\max_{1 \le p \le N}[(I_H(N-p) - I_H(N) - w_H(N)(-p)/p]\}.$$

By (4), we obtain

$$E(-\gamma_N^-) = E\{\max_{1 \le p \le N}[(I_H(-p)/p]\} .$$

Because $I_H(-p) \doteq I_H(p)$, we have $E(-\gamma_N^-) = E\gamma_N^+$. Therefore,

$$EF = 2E \max_{1 \le p \le N}[I_H(p)/p] = 2E \max_{1 \le p \le N} p^{-1}\int_0^p w_H(x)dx. \tag{10}$$

and $EF \le 2E \max[w_H(x), x \in (0,T)]$. Due to the self-similarity of H-FBM,

$$EF \le 2E \max[w_H(x), x \in (0,1)]T^H = 2M_1 T^H. \tag{11}$$

*The components of F: estimation.*

Expanding the range of p in (5, 6), one has

$$\gamma_k^- \ge \min_{1 \le p \le N}(I_H(k) - I_H(k-p))/p := \gamma_{k,N}^-,$$

$$\gamma_k^+ \leq \max_{1 \leq p \leq N}(I_H(k+p) - I_H(k))/p := \gamma_{k,N}^+,$$

whence

$$\gamma_k^- - \gamma_k^+ \geq \gamma_{k,N}^- - \gamma_{k,N}^+ \qquad (12)$$

By (4), we can continue

$$\doteq \gamma_{0,N}^- - \gamma_{0,N}^+ = \min_{-N \leq p \leq -1}[I_H(p)/p] - \max_{1 \leq p \leq N}[I_H(p)/p]. \qquad (13)$$

Using (12,13) and the increasing function $: x \to x_+$, we get

$$E(\gamma_k^- - \gamma_k^+)_+ \geq E(\min_{-N \leq p \leq -1}[I_H(p)/p] - \max_{1 \leq p \leq N}[I_H(p)/p])_+ := E\xi_N. \qquad (14)$$

By (11), (14),

$$2M_1 T^H \geq EF = E\sum_{k=1}^{N-1}[\gamma_k^- - \gamma_k^+]_+ \geq (T-2)E\xi_N.$$

Applying the Chebyshev inequality, one has

$$2M_1 T^H \geq (T-2)4P(\xi_N \geq 4). \qquad (15)$$

Because $\xi_N = (\gamma_{0,N}^- - \gamma_{0,N}^+)_+$,

$$P(\xi_N \geq 4) \geq P(\gamma_{0,N}^- \geq 2, \gamma_{0,N}^+ \leq -2) = P(I_H(p) + 2|p| \leq 0, |p| = 1,...,N)$$
$$\geq P(I_H(x) + 2|x| \leq 0, |x| \in (1,T)). \qquad (16)$$

Finally, for $T > 4$,

$$\widetilde{p}_T := P(I_H(x) + 2|x| \leq 0, |x| \in (1,T)) \leq M_1 T^{-(1-H)}. \qquad (17)$$

This relation is based on the following properties of the initial velocity:

affine invariance:

$$\{u_0(\lambda x + x_0) - u_0(x_0)\} \doteq \{|\lambda|^H u_0(x)\},$$

and integrability of the maximum:

$$M_1 = E\max[u_0(x), x \in (0,1)] < \infty.$$

Now we will use the Gaussian property of H-FBM to exclude the trend $2|x|$ in the persistence probability (17). We remind the following fact.

Let $\xi(x)$ be a centered Gaussian random process and $\varphi(x)$ is an element of the Hilbert space $H(\xi)$ with the reproducing kernel $E\xi(x)\xi(y)$ and the norm $\|\cdot\|_\xi$, (see e.g. [8]). Consider the persistence probabilities $p_T^{(\varphi)} := P\{\xi(x) + \varphi(x) \leq 1, x \in \Delta_T\}$. According to [1],

$$\left|\sqrt{-\log p_T^{(\varphi)}} - \sqrt{-\log p_T^{(0)}}\right| \leq \|\varphi\|_\xi / \sqrt{2}. \qquad (18)$$

To prove Proposition 3 using (18), we have to find $\varphi(x) \in H(I_H)$ such that

$$\varphi(x) = 2|x| + a, \qquad |x| \geq 1$$

with a=0 or a=1. By [11],

$$\psi(x) = 2x^2 \cdot 1_{|x|\leq 1} + (2|x|-1) \cdot 1_{|x|>1} \in H(I_H).$$

It only remains for us to find $\varphi(x) \in H(I_H)$ that is constant outside of (-1,1). Let $\eta$ be the unpredictable component of $I_H(1)$ given $\{w_H(x), x \in [0,1]^c\}$. Then $E\eta w_H(x) = 0, x \in [0,1]^c$. Therefore,

$$\varphi_1(x) = E\eta I_H(x) \in H(I_H), \varphi_1(x) = 0, x < 0, \varphi_1(x) = \varphi_1(1) = E\eta^2 = \|\varphi_1\|_{I_H}, x > 1.$$

Similarly we can find $\varphi_2(x) \in H(I_H): \varphi_2(x) = 0, x > 0$ and $\varphi_2(x) = const, x < -1$.

The desired function is $\varphi(x) = \psi(x) + (2+a)(\varphi_1(x)/\varphi_1(1) + \varphi_1(x)/\varphi_1(-1))$.

Finally, by (17, 18), one has

$$[\ln 1/P(I_H(t) < 1, t \in \Delta_T)]^{1/2} \geq \sqrt{\ln 1/\tilde{p}_T} - \|\varphi\|_{I_H}/\sqrt{2},$$

whence it follows that the inequality

$$\tilde{p}_T \leq M_1 T^{-(1-H)}$$

implies Proposition 3 for $a = 0$. If $a = 1$, we have to use the obvious inequality:

$$P\{I_H(x) \leq 1, x \in (-T,T)\} < P\{I_H(x) \leq 1, x \in (-T,T) \setminus (-1,1)\}.$$